\documentclass[10pt]{elsart}%
\usepackage{amssymb}
\usepackage{amsfonts}
\usepackage{amsmath}
\usepackage{graphicx}%
\begin{document}
%
\begin{frontmatter}%
%

\title
{The quantum-like description of the dynamics of party governance in the US political system}%
%

\author{Polina Khrennikova; Andrei Khrennikov; Emmanuel Haven}\footnote
{PK; EH: School of Management, University of Leicester, United Kingdom; AK: International Center for Mathematical Modelling in Physics and Cognitive Science, Linnaeus University, Sweden}%
%

\begin{abstract}%
This paper is devoted to the application of the mathematical formalism of
quantum mechanics to social (political) science. By using the quantum
dynamical equations we model the process of decision making in US elections.
The crucial point we attempt to make is that the voter's mental state can be
represented as a superposition of two possible choices for either republicans
or democrats. However, reality dictates a more complicated situation:
typically a voter participates in two elections, i.e. the congress and the
presidential elections. In both elections he/she has to decide between two
choices. This very feature of the US election system requires that the mental
state is represented by a 2-qubit state corresponding to the superposition of
4 different choices (e.g. for republicans in the congress; for the president
as a democrat). The main issue of this paper is to describe the dynamics of
the voters' mental states taking in account the mental and socio- political
environment. What is truly novel in this paper is that instead of using
Schr\"{o}dinger's equation to describe the dynamics in an absence of
interactions, we here apply the quantum master equation. This equation
describes quantum decoherence, i.e., resolution from superposition to a
definite choice.%

\end{abstract}%
%

\end{frontmatter}%

\thispagestyle{plain}

\section{Introduction}

For the last ten years, the mathematical formalism of quantum theory has been
actively applied outside the domain of quantum physics. We have seen numerous
applications in decision making (both in cognitive and social science),
economics and also finance. See for instance Acacio de Barros and Suppes
(2009) \cite{ABP09}, Asano et al. (2010) \cite{AOK}, Bruza et al. (2005
\cite{BC05}, 2009a \cite{BK09}, 2009b \cite{BWW09}); Busemeyer et al. (2006a
\cite{BMW06}, 2006b \cite{BWT06}); Cheon et al. (2006 \cite{CHEON}, 2010
\cite{CT10}); Choustova (2007 \cite{CHOUS}), Pothos et al. (2009) \cite{PB09},
Franco (2009) \cite{FF09}, Haven (2006 \cite{EH06}, 2008a \cite{EH08}, 2008b
\cite{EH08b}, 2009 \cite{EH09}) and La Mura (2008) \cite{LAM}.

Recently the \textit{quantum-like} (QL) approach started to be explored in
political science. Some of the QL features of the\textbf{\ }behavior of voters
in the US political system were discussed in Zorn and Smith (2011) \cite{ZSM}.
The authors start with a\textbf{\ }comparison of the notions of state
separability in conventional models of party governance and in quantum
information theory (see\textbf{\ }Zorn and Smith (2011) \cite{ZSM}) and they
then show that the QL model might provide a more adequate description of the
voters' state space -- `mental space'. The authors present a strong motivation
of the usage of the complex Hilbert space as the voters' `mental space.'

In this paper we present a QL-model describing the dynamics of the voters'
state (as represented in the complex Hilbert space). First, we consider what
we could call\textbf{\ }`a free QL-dynamics', when a voter\footnote{Following
the tradition of quantum information theory, we call such a voter `Alice'. In
game theory `Alice' is also often used.} is not under the pressure of mass
media and the social environment. By applying the quantum approach we describe
the dynamics of her state by using an analogue of the Schr\"{o}dinger
equation. A simple mathematical analysis implies that Alice's preferences
encoded in her state-vector (`mental wave function') fluctuate
\textit{without} stabilization to the definite state. Hence, such a dynamics
can describe the unstable part of the electorate: those voters who have no
firm preferences. In quantum physics, stabilization and damping of
fluctuations is a typical consequence of interaction with the
environment\footnote{Such environment in physics is also known as `bath'.}. We
apply this approach to the problem of the stabilization of fluctuations of
voters' preferences.

An essential part of the paper is devoted to the analysis of the applicability
of quantum dynamics to a social system (e.g. a voter) which is coupled to the
social environment. The main problem is that the exact quantum dynamics of a
system coupled to the physical environment is extremely complicated.
Therefore, to simplify matters, typically a quantum Markov approximation is
applied. This approximation is applicable under a number of non-trivial
conditions (see Ingarden et al. (1997) \cite{IKO}). Our aim is to translate
these conditions into the language of social science and to analyze their
applicability to the dynamics of voters' preferences. In this connection the
quantum Markovian dynamics, especially via the quantum master equation, can
model (approximately) voters' preference dynamics. Our approach is based on
the quantum master equation which describes the interaction of a social system
with a `social bath'. We use a very general framework which can be applied to
a variety of problems in politics, social science, economics, and finance. The
main problem of any concrete application is to analyze the conditions of
applicability of the quantum master equation (the quantum Markov
approximation) to the corresponding problem in decision making.

We remark that the work of Fiorina (1996) \cite{FI96} played an important role
in the motivation of the quantum model based on the use of entangled quantum
states (see Zorn and Smith (2011) \cite{ZSM} for the two institutional choices
in U.S. politics -- the congress and the presidency). Zorn and Smith (2011)
\cite{ZSM} also present a detailed analysis of the inter-relation between
classical and quantum models. Such an analysis is very important to attract
the interest of mainstream researchers in decision making to quantum models.
For such researchers, the applications of the quantum formalism to social
science may on prima facie be considered as quite exotic. Therefore, in this
paper, we begin with an extended section in which we compare classical and
quantum probabilistic approaches to decision making. Our aim is not only to
stress the differences, but also to find the commonalities. Our findings argue
for an important degree of similarity between quantum and subadditive
probabilistic descriptions of decision making. We also emphasize the vital
role of contextuality.

\section{Inter-relation between quantum and traditional models of decision
making}

\subsection{(Non-) Bayesian approach}

One of the basic tools of probabilistic investigations in psychology,
cognitive science, economics and finance is Bayesian analysis (see\textbf{\ }%
De Finetti (1972) \cite{DF72} and Kreps (1988) \cite{KREP}) which allows for a
process of mental updating of probabilities (objective or subjective depending
on the interpretation\footnote{There are two `camps' around the
interpretations of Bayesian probabilities divided in so called `objectivists'
and `subjectivists'. See Chalmers (1999) \cite{CH99}.}) on the basis of newly
collected statistical data. Bayesian probability can be distinguished to be
objective (independent of the individual who makes a decision) or subjective,
that is to say, related to the\textbf{\ }personal belief of an individual
(see\textbf{\ }De Finetti (1974) \cite{DF72}). The objective probabilities
represent the choice that rational agents should make in the light of an
objective situation and updating occurs as a consequence of the appearance of
any new event (see Chalmers (1999) \cite{CH99}). By this approach the agents
are supposed to distribute the prior probabilities equally on the basis of
some principle of indifference. In particular, the Bayesian approach plays an
important role in classical decision making (see De Finetti (1972)
\cite{DF72}). We stress that this method is a part of conventional
(`classical') probability theory based on Kolmogorov's axiomatics (1950)
\cite{KOL}. The Bayes formula for conditional probabilities is:%
\begin{equation}
P(A|B)=P(AB)/P(B); \label{1}%
\end{equation}
where $P(B)\neq0.$ The law of total probability forms an integral part of the
classical Bayesian approach. Let us consider the law of total probability in
the simplest situation. Consider an event $B$ and its complement $\bar{B}$ and
assume that the probabilities of both these events are positive. Then, for any
event $A,$ the following formula (of total probability) holds:
\begin{equation}
P(A)=P(B)P(A|B)+P(\bar{B})P(A|\bar{B}). \label{2}%
\end{equation}
We note that for \textit{quantum probabilities}, the law of total probability
is violated! In general, the difference between the left-hand and right-hand
sides of (2) is nonzero. This difference is nothing else than the influence of
the interference term, which plays a fundamental role in quantum theory (as
well as in classical physical wave theories). The quantum analog of the law of
total probability has the form:
\begin{equation}
P(A)=P(B)P(A|B)+P(\bar{B})P(A|\bar{B})+2cos\;\theta\sqrt{P(B)P(A|B)P(\bar
{B})P(A|\bar{B})}. \label{3}%
\end{equation}
Depending on the sign of $cos\;\theta$ one observes constructive
($cos\;\theta>0)$ or destructive ($cos\;\theta<0)$ interference. In the first
case the probability to observe some phenomenon increases so much that it
cannot be explained by the laws of classical probability theory. In the second
case one similarly finds a `mystical' decreasing of probability (e.g.,
probabilities $P_{1}=P_{2}=1/2$ can result in a zero probability, $P_{12}=0).$
In the case $cos\;\theta=0$ the quantum formula of total probability (the
formula containing thus an interference of probabilities) is reduced to the
classical law of total probability. This is a very important point of
transition from usage of the classical probabilistic model to the quantum
probabilistic model.

By decreasing the absolute value of interference coefficient, the latter can
be transformed into the former (as the coefficient vanishes). Thus the quantum
probabilistic models in cognitive science, psychology, and social science are
natural extensions of the classical models. If the deviation of the left-hand
side of equation (3) from the right-hand side is relatively small, we can
ignore the interference contribution and proceed with the classical law of
total probability.

In summary, we can think of the quantum approach of decision making as a
natural generalization of the Bayesian approach which is based on the
transition from the classical formula of total probability to its quantum analogue.

\subsection{Subadditive probability in social science, psychology, behavioral
economics and finance}

The following questions naturally arise when considering the use of equation
(3) as a new tool in decision making:

\begin{enumerate}
\item Is the departure from equation (2) to equation (3) a totally new step in
the development of probabilistic modeling in social science?

\item Are there other conventional social models based on departures from the
laws of classical probability?
\end{enumerate}

Surprisingly for those who argue for the exceptional novelty of the quantum
approach to social problems, the answer is `yes'. In mainstream studies in
cognitive science, psychology, behavioral economics and finance, non-classical
probability has been actively used during many years.

Comparing the quantum approach with the traditional non-classical
probabilistic approaches is not a straightforward task. In QL models the
\textit{violation of the law of total probability} is considered as the
crucial point. However, the majority of traditional non-classical models are
not based on the aforementioned violation of equation(2), but rather on
an\textbf{\ }application of \textit{subadditive probabilities.} Hence,
the\textbf{\ }\textit{violation of the law of additive probability} has
already been\textbf{\ }actively discussed in social science. Khrennikov and
Haven (2007) \cite{KH07} indicate (p. 23) that\textbf{\ \textquotedblleft}when
experiment participants have to express their degree of beliefs on a [0, 1]
interval, probabilistic additivity will be violated in many cases and
subadditivity obtains. See Bearden et al. (2005) \cite{BWF} for a good
overview.\textquotedblright\ Khrennikov and Haven (2007) \cite{KH07} continue
as follows (p. 23-24): \textquotedblleft Bearden et al. (2005) \cite{BWF} also
indicate that such subadditivity has been obtained with experiment
participants belonging to various industry groups, such as option traders for
instance (Fox et al. (1996) \cite{FRT96}). The key work pertaining to the
issue of subadditivity in psychology is by\textbf{\ }Tversky and Koehler
(1994) \cite{TK94} and Rottenstreich and Tversky (1997) \cite{RT97}. Their
theory, also known under the name of `Support Theory' is in the words of
Tversky and Koehler (1994) \cite{TK94} \ `...a theory in which the judged
probability of an event depends on the explicitness of its description.' In
other words, it is not the event which is important as such but its
description. In Tversky and Koehler (1994) \cite{TK94} the authors highlight
the `current state of affairs'...on the various interpretations that
subjective probability may have. Amongst the interpretations is Zadeh's (1978)
\cite{ZAD} possibility theory and the upper and lower probability approach of
Suppes (1974) \cite{SUPP}. The paper of Dubois and Prade (1998) \cite{DP88},
also mentioned in Tversky and Koehler (1994) \cite{TK94}, provides for an
excellent overview on non-additive probability approaches.\textquotedblright\ 

To couple QL models based on the violation of the law of total probability and
Bayesian probability, with traditional studies based on subadditive
probabilities, we need to recall that the mathematical derivation of the
formula of total probability is based on the additivity of probability and the
Bayes formula for conditional probabilities. Therefore there are two possible
sources of violation of equation (2): i) subadditivity and ii) the
non-Bayesian definition of conditional probability. Both these sources exhibit
themselves in QL-models. Hence, the subadditivity of probability is an
important common point of the QL and traditional (based on non-classical
probability) approaches. Moreover, many experts in quantum physics especially
stress the role of subadditivity of quantum probability as the main source of
quantum interference (Feynman and Hibbs (1965) \cite{FH65}.

The QL approach can be considered as a special mathematical model describing
the usage of subadditive probability in social science. It is not clear
whether any social science based model with subadditive probability can be
embedded in the QL-approach. The quantum probabilities have a very special
structure: they are based on complex probability amplitudes, vectors from a
complex linear space and probability is obtained from a squared complex
amplitude. \textit{It is not clear whether any subadditive probability from
the aforementioned social science based} \textit{models can be represented in
this way.} Nevertheless, even if it might imply the loss of generality, the
use of the linear space representation simplifies the operation with
probabilities. Furthermore, it provides us with a possibility to use a
powerful mathematical apparatus of quantum mechanics in an interdisciplinary way.

In this paper, we intend to explore quantum dynamical equations. We stress
that the form of these equations depend very much on whether interaction with
the environment is taken into account or neglected. Here we are merely
interested in the application of quantum dynamics to the modeling of the
evolution of the mental state of a human being interacting with an extremely
complex social environment. The complexity of the actual environment is so
high that it strongly influences the decision making process of an individual,
finally implying a resolution from superposition of his/her mental states. In
quantum terms a decoherence takes place.

\subsection{Savage sure thing principle and disjunction effect}

Bayesian probability according to Maher (2010) \cite{MAH} \ (p. 120)
\textquotedblleft explicates a kind of rationality we would like our choices
to have...\textquotedblright\ Correspondingly, the `absolute rational choice'
Maher (2010) (p.120) refers to, can be understood as \textquotedblleft the
maximization of expected utility.\textquotedblright\ The Bayesian updating of
probabilities and the validity of the law of total probability have a direct
coupling with the problem of rationality in decision making. von Neumann and
Morgenstern's (1944) \cite{VNM44} expected utility theory, and Savage's sure
thing principle (Savage (1954) \cite{SAV}) postulate a complete rationality
(i.e. a maximization of one's own payoff and the minimizing of one's own
losses). Savage (1954) \cite{SAV} (p. 21) proposed the so called
\textbf{`}Sure thing principle (STP)', denoting that: \textquotedblleft If a
person would not prefer [a decision] $f$ to $g,$ either knowing that the event
$B$ obtained, or knowing that the event $\bar{B}$ obtained, then he does not
prefer $f$ to $g$ [whether knowing or not if the event $B$ or $\bar{B}$
happened].\textquotedblright\ Savage (1954) \cite{SAV}\textbf{\ }%
illustrate\textbf{s} the validity of the principle with an example of a
businessman, who considers whether to buy some property before the
presidential elections or not. Savage (1954) (p. 21) describes the situation
of a businessman who is uncertain if the Republicans or Democrats will win the
election campaign. He decides that he would buy the property if the
Republicans win, but also he decides that he should buy the
property\textbf{\ }even if the Democrats win. By taking the decision to buy in
any case (for example the decision $g$), it is natural to assume that the
businessman will buy the property being uncertain of whether the Republicans
(event $B)$ or Democrats will win (event $\bar{B})$. The principle could be
statistically represented with help of the formula of total probability
(equation (2)), where the events $B$ and $\bar{B}$ are assigned some
probability and the decision $A$ (here depicted as $g,$ to be consistent with
Savage's symbols) would be a conditional probability of $B$ and $\bar{B},$ so
that the exact statistical probability for the possible decision $g$ could be
obtained. In this illustration we see that the conditional probability of $g$
would be equal to one (i.e. there is $100\%$ confidence about the purchase of
a house). According to Croson (1999) \cite{CROS} the event $B$ can be as well
i)\textbf{\ }an exogenous risk: the uncertainty about the state of nature
(e.g. the property purchase) as well as ii)\textbf{\ }a strategic risk: an
uncertainty about the choice of a strategic opponent (e.g. a competitor starts
a price war). Croson (1999) \cite{CROS} describes such a pattern of decision
making as `consequential reasoning', as the individual considers the
consequences (for instance the amount of the payoffs) before considering a
particular action.

Savage's Sure thing Principle has been regarded as a foundation axiom for
decision making in economics. Kreps (1998) \cite{KREP} (p. 120) called
Savage's principle the \textquotedblleft crowning glory of choice
theories\textquotedblright. However, many experiments, such as Allais (1953)
\cite{AM53}, Tversky and Shafir (1992) \cite{TS92}, Croson (1999) \cite{CROS}
proved that economic decision makers in general tend to violate the Savage
sure thing principle and expected utility theory. For example, in a Prisoner
Dilemma type game experiment, violation of the rationality postulate of
Savage's sure thing principle was found in experiments performed by Tversky
and Shafir (1992) \cite{TS92} and later repeated by Croson (1999) \cite{CROS}
and Busemeyer et al. (2006) \cite{BMW06}. Traditionally, this game is played
in three conditions. In the `unknown' condition the player acts without
knowing the opponent's action. In the known `defect condition', the player
knows that the opponent has defected before he/she acted. In the known
`cooperate condition' the player knows that the\textbf{\ }opponent has
cooperated, before he/she acted. See also Tversky and Shafir (1992)
\cite{TS92} and Pothos and Busemeyer (2009) \cite{PB09}.

We cite Tversky and Shafir (1992) \cite{TS92} (p. 309) \textquotedblleft The
subjects... played a series of prisoners dilemma games, without feedback, each
against a different unknown opponent supposedly selected at random from among
the participants. In this setup the rate of cooperation was 3\% when subjects
knew that the subject knew that the opponent has defected and 16\% when they
knew that the opponent has cooperated. However, when the subjects did not know
whether their opponent had cooperated or defected (as is normally the case of
the game) [condition of uncertainty]) the rate of cooperation rose to
37\%.\textquotedblright\ This experiment showed that when the players are
unaware of their opponents actions, they do not behave rationally as they are
supposed to do in a conventional prisoners dilemma game. This anomaly in
behavior occurred in other games of the Prisoners Dilemma type and also in
Hawaiian vacation experiments. The basic effect those experiments have in
common is referred to by Tversky and Shafir (1992) \cite{TS92} and Croson
(1999) \cite{CROS} as the `disjunction effect'. Busemeyer et al. (2006)
\cite{BMW} show that the disjunction effect is equivalent to the\textbf{\ }%
violation of the law of total probability. Since this law is violated by QL
models, all such models in social science exhibit the disjunction affect.

\subsection{Contextuality}

In the quantum community there is still no consensus on the basic roots of
`quantum mysteries'; in particular, the grounds for the violation of the laws
of classical probability theory. One hundred years after the creation of
quantum mechanics (the 1920's-1930's starting with the founders of quantum
mechanics : Bohr, Heisenberg and Einstein), the intensity of debates about its
foundation have not abated. We may even claim the debates are more intense.
One of the possible sources of the quantum mysteries is the notion of
contextuality. The viewpoint that the results of quantum observations depend
crucially on the measurement context was proposed by Niels Bohr, who
emphasized that we are not able to approach the micro world (with the aid of
our measurement devices) without bringing essential disturbances into its
state. The quantum systems are too sensitive to the measurement apparata. The
context of measurement plays an essential (depending on the interpretation,
even crucial) role in forming the result of our measurement. According to the
fundamental interpretation of quantum mechanics, the `Copenhagen
interpretation', quantum systems do not have objective properties which exist
independently of `questions' asked to these systems in the context of
measurement. Says Suppes (1974) \cite{SUPP} (p. 171-172)\textbf{:
\textquotedblleft}Any time we measure a microscopic object by using
macroscopic apparatus we disturb the state of the microscopic object and,
according to the fundamental ideas of quantum mechanics, we cannot hope to
improve the situation by using new methods of measurement that will lead to
exact results of the classical sort for simultaneously measured conjugate
variables.\textquotedblright

The contextual viewpoint is attributed to the origin of non-classical
probabilistic behavior of quantum systems and is very attractive for those who
already apply or aim to apply a quantum formalism in other domains outside
physics. It is important to stress that the contextual interpretation of
quantum mechanics is more `innocent' than other essentially more exotic
viewpoints, such as the quantum non- locality concept or the `many worlds'
interpretations. The majority of people working in cognitive science and
psychology would not accept a possibility of non-local interactions between
human beings, e.g. through a splitting of reality in many worlds.

The concept of contextuality is a well known feature of cognitive systems. We
also see the origin of non-Bayesian (`irrational'\footnote{For a comparison,
please see section 2.3. (especially Maher (2010) \cite{MAH})}) decision making
in the contextuality of observations performed for mental quantities,
including self-observations. Hence, the value of the subjective probability
does not exist independently of the measurement context, only whilst `asking
about someone's preferences' including ourselves, we create them.

For example, in semantics studies context is treated by representing it as cue
words, or co- appearing words. This semantic contextuality (well known and
actively explored in traditional semantic models) was used as the starting
point for the development of QL model\textbf{s} of word recognition (see Bruza
et al. (2005 \cite{BC05}, 2009a \cite{BK09}, 2009b \cite{BWW09}). We also
remark that contextual models of reasoning play an important role in
artificial intelligence (see f.i. Giunchiglia (1993) \cite{G193}, McCarthy
(1993) \cite{MCC}.

We now come back to the problem of rationality in decision making. We remark
that contextuality of reasoning is closely coupled with the so called `framing
effect'. Kreps (1988) \cite{KREP} remarks (p. 197) that \textquotedblleft the
way in which a decision problem is framed or posed can affect the choices made
by decision makers.\textquotedblright\ According to Tversky and Kahnemann
(1981) \cite{TK81} the term `decision frame' refers (p. 453) \textquotedblleft
to the decision-maker's conception of the acts, outcomes and contingencies
associated with a particular choice.\textquotedblright\ One of the most
important contributions of the QL approach to the problem of contextual
reasoning is the recognition of the existence of incompatible contexts and the
use of well developed quantum tools for testing incompatibility, such as
Heisenberg's uncertainty relation or the violation of Bell's inequality (see
f.i. Khrennikov and Haven (2007) \cite{KH07}). In particular, in
the\textbf{\ }Prisoner's Dilemma game the contexts $C_{\mathrm{known}}$ (the
decision of the partner is known), and $C_{\mathrm{unknown}}$ (information is
absent), are incompatible. Consequently, the QL approach is about:

\begin{enumerate}
\item the violation of the Sure Thing Principle,

\item `irrational' decision making,

\item non-Bayesian decision making, and

\item the usage of subadditive probability.
\end{enumerate}

All these problems have already been widely discussed in traditional
approaches to cognitive science, psychology, behavioral economics and finance.
The QL approach is just one of the mathematical models which accurately
describes all of the above effects.

Finally, we point to one of the pioneer papers that assigned quantum-like
contextuality to the measurement of belief in decision making theories. Suppes
(1974) \cite{SUPP} conjectured that general concepts taken from quantum
mechanics could provide for the measurement of belief. He also explained the
importance of the particular measurement context, by asserting that (p. 172):
`it is a mistake to think of beliefs as being stored in some fixed and inert
form in the memory of a person. When a question is asked about personal
beliefs, one constructs a belief coded in a belief statement as a response to
the question. As the kind of question varies, the construction varies, and the
results vary.\textquotedblright\ 

\textbf{W}e could articulate that the notion of measurement context, borrowed
from quantum mechanics can be regarded as one of the promising theories of
measurement of belief.

\section{Description of election campaign by the theory of open quantum
systems}

With the help of the above mentioned features of QL models we now attempt to
describe the dynamics of the process of decision making within the problem
setting of party governance in the US-type two party system. This system
allows voters to cast partisan ballots in two contests: executive and
legislative. By so doing they can thus choose for instance `Republican' in one
institutional choice setting and `Democratic' in the other (see Zorn and Smith
(2011) \cite{ZSM}).

It is well known from physics that the quantum state dynamics are described by
Schr\"{o}dinger's equation. This type of dynamics is unitary. Roughly speaking
it is combined of a family of rotations and in principle, this family can be
infinite. Pothos and Busemeyer (2009) \cite{PB09} applied this equation to
model the dynamics of the process of decision making in games of the
Prisoner's Dilemma type. However, it is questionable whether we can describe
the dynamics of voters' expectation by the Schr\"{o}dinger's equation. This
equation describes the dynamics of an isolated system, i.e., a system which
does not interact with the environment. A voter in the context of the election
campaign definitely cannot be considered as an isolated social system. She,
say Alice, is in permanent contact with mass media (whether TV or internet).
Such an influence of the environment induces random fluctuations of opinions
and choices in Alice's mind.

For the purposes of our research, we are interested in the `unstable' part of
the electorate which is composed of citizens who have no concrete opinions and
who will make their electoral choice\ very close to the actual day of the
elections (see Zaller and Feldman (1992) \cite{ZF92}).

If Alice could be considered as an isolated social system, then the only
possibility to describe a transition from the mental state of superposition of
choices to the state corresponding to the concrete choice was to use the
projection postulate of quantum mechanics (the so called `von Neumann
postulate'). This state reduction process, from superposition to one of its
components, is called \textit{the state collapse}\footnote{The state collapse
is considered as one of the main mysteries of quantum physics. This notion is
still a subject of intensive debate.}. Such collapse is imagined as an
instantaneous (the jump-type) transition from one state to another. The state
collapse might be used to describe the situation in which Alice makes her
choice precisely at the moment of completing the voting bulletin. This type of
behavior cannot be completely excluded from consideration, but such a case is
probably not statistically significant. Moreover, mainstream quantum
mechanical thought will tell us that the state collapse occurs when an
isolated system driven by Schr\"{o}dinger's equation interacts practically
instantaneously with a measurement device. Thus when Alice is totally isolated
from the election campaign, she is suddenly asked to make her choice. It is
evident that the process of decision making for the majority of the `unstable
population' in the electorate differs in essential ways from this
collapse-type behavior.

Therefore, let us take more seriously the role which the social environment
plays in the process of decision making. We apply to social science the theory
of \textit{open quantum systems}, i.e., systems which interact with a large
thermostat (`bath'). Since a bath is a huge physical system with millions of
variables (the complexity of the \textquotedblleft social
bath\textquotedblright\ around an American citizen who will cast his/her vote
in the election campaign is huge), it is in general impossible to provide a
reasonable mathematical description of the dynamics of a quantum system
interacting with such a bath. Physicists proceed under a few assumptions which
allow then for the possibility to describe those dynamics in an approximate
way. In quantum physics the interaction of a quantum system with a bath is
described by a quantum version of the master equation for Markovian dynamics.
The quantum Markovian dynamics are given by the
\textit{Gorini-Kossakowski-Sudarshan-Lindblad} (GKSL) equation. See e.g.
Ingarden et al. (1997) \cite{IKO} for details. This GSKL equation\textbf{\ }is
the most popular approximation of quantum dynamics in the presence of
interaction with a bath.

We briefly\textbf{\ }remind the origin\textbf{s} of the GKSL-dynamics. The
starting point is that the state of a composite system, a quantum system $s$
combined with a bath, is a pure quantum state, complex vector $\Psi.$ The
evolution of\textbf{\ }$\Psi$ is described by Schr\"{o}dinger's equation. This
is an evolution in a Hilbert space of a huge dimension, since a bath has so
many degrees of freedom. The existence of the Schr\"{o}dinger dynamics in the
huge Hilbert space has a merely theoretical value. Observers are interested in
the dynamics of the state $\phi_{s}$ of the quantum system $s.$ The next
fundamental assumption in the derivation of the GKSL-equation is the Markovian
character of the evolution, i.e. the absence of long term memory effects. It
is assumed that interaction with the bath destroys such effects. Thus, the
GKSL-evolution is a Markovian evolution. Finally, we point to the condition of
the `factorizability' of the initial state of a composite system (a quantum
system coupled with a bath), $\Psi=\phi_{s}\otimes\phi_{\mathrm{bath}},$ where
$\otimes$ is the sign of the tensor product. Physically factorization is
equivalent to the absence of correlations\footnote{At the beginning of
evolution; later they are induced by the interaction term of Hamiltonian --
the generator of evolution.}. One of the distinguishing features of the
evolution under the mentioned assumptions is the existence of one or few
\textit{equilibrium points.} The state of the quantum system $s$ stabilizes to
one of such points in the process of evolution: a pure initial state, a
complex vector $\psi_{s},$ is transformed into a mixed state, a density matrix
$\rho_{s}(t)$ (classical state without superposition effects).

In contrast to the GKSL-evolution, the Schr\"{o}dinger evolution does
\textit{not} induce stabilization. Any solution different from an eigenvector
of the Hamiltonian will oscillate for ever. Another property of the
Schr\"{o}dinger dynamics is that it \textit{always} transfers a pure state
into a pure state, i.e., a vector into a vector: quantumness if it was
originally present in a state (in the form of superposition) cannot disappear
in the process of a continuous dynamical evolution. The transition from
quantum indeterminism to classical determinism can happen only as the result
of the collapse of the quantum state.

On the one hand, in our model of the decision making for party governance we
would like to avoid the usage of the state collapse. On the other hand, to
make a decision, Alice has to make a transition from a quantum to a classical
representation of her preferences. We note that in quantum physics all
experimentally obtained information is classical as well. The GKSL-evolution
provides for such a possibility (and without `quantum jumps'). Alice's mental
state evolves in a\textbf{\ }smooth way (fluctuations exist but they are
damped) to the final classical decision state.

\section{Matching of assumptions of applicability of the quantum master
equation with conditions of the modern election campaign}

We now list the social conditions corresponding to the above mentioned
physical conditions. This will allow us for a possibility to apply the GKSL-equation:

\begin{itemize}
\item (\textbf{COMPL})\textit{\ complexity:} the social environment (election
bath) influencing a voter has huge complexity

\item (\textbf{FREE}) \textit{freedom: }the mental state of a society under
consideration is a pure QL state, i.e., \textbf{a }superposition of various
opinions and expectations;

\item (\textbf{DEM}) \textit{democracy}: the feedback reaction of a voter to
the election bath is negligibly small, it cannot essentially change the mental
state of the bath;

\item (\textbf{SEP}) \textit{separability}: before the start of the election
campaign a voter was independent of the election bath;

\item (\textbf{MARK}) \textit{Markovness}: a voter does not use a long range
memory on interaction with the election bath to update her state.
\end{itemize}

We surely need to make some comments on those assumptions.

\begin{enumerate}
\item The assumption (\textbf{COMPL}), \textit{complexity}, is definitely
justified. Nowadays an election campaign has huge information complexity: the
richness of media sources accounts for such complexity. We can even speculate
that the proposed QL model is more adequate than say 50 years ago: the
phenomenal increase of information complexity makes the usage of the (quantum,
quantum-like) open systems approach more reasonable.

\item The (\textbf{FREE}), \textit{freedom}, can be interpreted as
guaranteeing the freedom of political opinions. The opposite to the
(\textbf{FREE})-society, is a totalitarian society where its mental state is a
classical state in which all superpositions have been resolved (collapsed).

\item The (\textbf{DEM}), \textit{democracy}, encodes the democratic system:
one voter cannot change the mental state of society in a crucial way.

\item The (\textbf{SEP}), \textit{separability}, describes a sample of voters
who are not that interested in politics: they will determine their positions
through an interaction with the election bath during the election campaign.
This part of the electorate is the most interesting from the point of view of
political technologies.

\item The (\textbf{MARK})-assumption, \textit{Markovness}, also reflects the
fact that voters under study are not that interested in politics. They do not
spend a lot of time analyzing the dynamics of the election campaign. However,
they are not isolated from the election bath; they watch TV, read newspapers
and use the internet. From a pragmatic point of view, they unconsciously
update their mental states each day by taking into account recent news.
\end{enumerate}

\begin{rem}
[Markovness]We remark that the Markovness of the dynamics may induce the
impression that voter's preferences would fluctuate forever. However, this is
not the case. The mathematical formalism of quantum mechanics implies that
quantum Markovean fluctuations stabilize to steady solutions. In physics, this
theoretical prediction was confirmed by numerous experiments. Although the
social counterparts of physical assumptions seem to be natural and this
motivates the applicability of our theoretical model, the final justification
can come only from the testing of our hypothesis by experimental data. This is
a very complex problem.
\end{rem}

\begin{rem}
[Decoherence]In quantum physics the process of transformation of a pure
(superposition-type) state into a classical state (given by a diagonal density
matrix) is called decoherence. A proper interpretation of this process is
still one of the hardest problems \textit{in the }foundations of quantum
mechanics. Some authors present the viewpoint that superposition is in some
way conserved: the disappearance of superposition in a subsystem increases it
in the total system. In our model this would mean that the determination of
states of voters in the process of interaction with the election bath will
\textit{transfer political uncertainty into an increase of political
uncertainty in society in general, after elections.} At the moment it is not
clear whether this interpretation is meaningful in social sciences.
\end{rem}

\section{Schr\"odinger's dynamics}

The state space of a voter (Alice) can be represented as the tensor product of
two Hilbert spaces (each of them is two dimensional). One Hilbert space
describes the election to the congress, and we denote it by the symbol
$\mathcal{H}_{\mathrm{congress}},$ and another describes the presidential
election, denote it by the symbol $H_{\mathrm{president}}.$ In each of them we
can select the basis corresponding to the definite strategies $e_{1}%
=|d\rangle,e_{2}=|r\rangle.$ If Alice was thinking only about the election to
congress, her mental state would be represented as the superposition of these
two basis vectors:
\begin{equation}
\psi_{\mathrm{congress}}=\alpha_{c}|d\rangle+\beta_{c}|r\rangle; \label{CONG}%
\end{equation}
where $\alpha_{c},\beta_{c}$ are complex numbers and they are normalized by
the condition: $|\alpha_{c}|^{2}+|\beta_{c}|^{2}=1.$ By knowing the
representation of equation (\ref{CONG}) one can find the probabilities of
intentions to vote for democrats and republicans in the election to the
congress:
\begin{equation}
p_{\mathrm{congress}}(d)=|\alpha_{c}|^{2},\;p_{\mathrm{congress}}%
(r)=|\beta_{c}|^{2}. \label{CONG1}%
\end{equation}
However, the quantum dynamics of the state, $\psi_{\mathrm{congress}}(t),$ in
the absence of interactions with the political bath (environment), see
equation (\ref{SCH1}) below -- `social Schr\"{o}dinger equation', is such that
the probabilities $p_{\mathrm{congress}}(d;t),$\newline$p_{\mathrm{congress}%
}(r;t)$ fluctuate. Therefore, even if Alice wanted to vote for republicans at
$t=t_{0},$ in the process of mental evolution she will change her mind many times.

In the same way, if Alice w\textbf{as} thinking only about the election of the
president, her mental state would be represented as \textbf{a }superposition
of the two basis vectors
\begin{equation}
\psi_{\mathrm{president}}=\alpha_{p}|d\rangle+\beta_{p}|r\rangle;
\label{CONG2}%
\end{equation}
where $|\alpha_{p}|^{2}+|\beta_{p}|^{2}=1.$ The corresponding probabilities
are given by
\begin{equation}
p_{\mathrm{president}}(d)=|\alpha_{p}|^{2},\;p_{\mathrm{president}}%
(r)=|\beta_{p}|^{2}. \label{CONG1}%
\end{equation}
For a moment, let us forget about the quantum model and turn to classical
probability theory. Suppose that the classical probabilities
$p_{\mathrm{congress}}(d)$, $p_{\mathrm{congress}}(r)$, $p_{\mathrm{president}%
}(d)$, $p_{\mathrm{president}}(r)$ are given. Furthermore, suppose that voters
do not have any kind of correlations between two elections: their choice in
the election to the congress does not depend on their choice of the president
and vice versa\footnote{To escape the problem of time fluctuations of
probabilities, we may assume that both elections are done at the same time.}.
In this case independence implies factorization of the joint probability
distribution:
\begin{equation}
p_{\mathrm{{congress},{president}}}(dd)=p_{\mathrm{congress}}%
(d)p_{\mathrm{president}}(d),... \label{CONG2_K}%
\end{equation}
However, in the case of non-trivial correlations between the congress- and
president-elections, the factorization condition is violated. In the quantum
formalism, the models described by the two Hilbert spaces are unified in the
model described by the tensor product of these two spaces. In our case we use
the space $\mathcal{H}=\mathcal{H}_{\mathrm{congress}}\otimes
H_{\mathrm{president}}.$ Its elements are of the form:
\begin{equation}
\psi=\psi_{\mathrm{congress}}\otimes\psi_{\mathrm{president}} \label{CONG2_K1}%
\end{equation}
which describe the states corresponding to uncorrelated choices in two
elections. In quantum information such states are called \textit{separable.}
In general a state $\psi\in\mathcal{H}$ cannot be factorized. Nonseparable
states describe correlations between choices in the two elections:
\begin{equation}
\psi=c_{dd}|dd\rangle+c_{dp}|dp\rangle+c_{pd}|pd\rangle+c_{pp}|pp\rangle;
\label{CONG2_K2}%
\end{equation}
where $|c_{dd}|^{2}+|c_{dp}|^{2}+|c_{pd}|^{2}+|c_{pp}|^{2}=1$ and
$|dd\rangle=|d\rangle\otimes|d\rangle,....$ The main point of usage of the
quantum formalism is that quantum correlations are not reduced to classical
correlations (as\textbf{\ }described in the framework of the Kolmogorov
model). Roughly speaking the quantum correlations can be stronger than the
classical correlations. This is the essence of Bell's theorem, Bell (1987)
\cite{B87}. We also state again that the question of inter-relation between
quantum and classical separability in the election framework was studied in
Zorn and Smith (2011) \cite{ZSM}. However, the authors\textbf{\ }did not
appeal directly to Bell's theorem, but to the more delicate condition of
quantum (non-) separability. Its role in social science was emphasized by
Bruza et al. (2010) \cite{BRUZA}.

The quantum dynamical equation has the form:
\begin{equation}
ih\frac{\partial\psi}{\partial t}(t)=H\psi(t); \label{SCH}%
\end{equation}
where $H$ is the operator of energy, the Hamiltonian, and $h$ is the Planck
constant. The mental interpretation of an analog of the Planck constant $h$ is
a complicated problem. We shall interpret it as the time scale
parameter\footnote{In quantum physics the Planck constant has the dimension of
action: $\mathrm{{time}\times{energy}.}$ In this paper we do not want to
speculate on such a controversial topic as \textquotedblleft mental
energy\textquotedblright\ (but see however, Choustova (2007) \cite{CHOUS}).
Therefore, we proceed formally by considering the evolution generator $H$ as a
dimensionless quantity.}. Since the usage of the symbol $h$ may be a source of
misunderstanding (especially for physical science educated readers), we shall
use a new scaling parameter, say $\tau$ having the dimension of time (please
see the preceding footnote). It determines the time scale of updating of the
mental state of Alice during the election campaign. We rewrite the dynamical
equation as:%
\begin{equation}
i\tau\frac{\partial\psi}{\partial t}(t)=H\psi(t). \label{SCH1}%
\end{equation}
And we call the operator $H$ , the\textbf{\ }\textit{decision Hamiltonian.}

The most general Hamiltonian $H$ in the space of mental states in the
two-party systems (wherein voters can cast partisan ballots in two contests,
executive and legislative) has the form
\begin{equation}
H=H_{\mathrm{stab}}+H_{\mathrm{flip}}, \label{HAM}%
\end{equation}
where $H_{\mathrm{stab}}$ is the part of the Hamiltonian responsible for the
stability of the distribution of opinions about various possible selections of
decisions. It is given by
\begin{equation}
H_{\mathrm{stab}}=\lambda_{dd,dd}|dd\rangle\langle dd|+\lambda_{rr,rr}%
|rr\rangle\langle rr|+\lambda_{dr,dr}|dr\rangle\langle dr|+\lambda
_{rd,rd}|rd\rangle\langle rd|. \label{HAM1}%
\end{equation}
And $H_{\mathrm{flip}}$ is the part of Hamiltonian responsible for flipping
from one selection of the pair of strategies (for executive and legislative
branches) to another. It is given by
\begin{equation}
H_{\mathrm{flip}}=\lambda_{dd,rr}|dd\rangle\langle rr|+\lambda_{rr,dd}%
|rr\rangle\langle dd|+\lambda_{rd,dr}|rd\rangle\langle dr|+\lambda
_{rd,dr}|rd\rangle\langle dr|. \label{HAM12}%
\end{equation}
To induce a unitary evolution, the Hamiltonian has to be Hermitian. This
induces the following restrictions to its coefficients: $\lambda
_{dd,dd},...,\lambda_{rd,rd}$ are real and $\overline{\lambda}_{dd,rr}%
=\lambda_{rr,dd},...,\overline{\lambda}_{rd,dr}=\lambda_{dr,rd}.$

In the absence of the $H_{\mathrm{flip}}$-component, the probabilistic
structure of superposition is preserved. Only phases of choices evolve in the
rotation-like way, e.g., $|dd\rangle$ evolves as
\[
\psi_{dd}(t)=e^{-\frac{it}{\tau}\lambda_{dd,dd}}|dd\rangle
\]
which corresponds to \textbf{a }\textquotedblleft rotation" of the strategy
$dd$ for the \textquotedblleft angle\textquotedblright\ $\Delta\theta
=t\lambda_{dd,dd}.$ \textbf{A }larger $\lambda$ induces quicker rotation. The
meaning of such rotations of mental states has to be clarified in the process
of the model's development. We can speculate that the coefficient
$\lambda_{dd,dd}$ correspond to the speed of self-analysis (by Alice) of the
choice $dd.$

In the presence of the flipping component $H_{\mathrm{flip}}$ the distribution
of probabilities of choices of various strategies changes in the process of
evolution. Such flipping from one strategy to another makes the state dynamics
really quantum. In fact, for political technologies per s\'{e}, the most
important component is the flipping part of the Hamiltonian. Of course, at the
moment we proceed at \textbf{a} very abstract theoretical level. However, one
may hope to develop the present QL model to the level of real applications.

\begin{exmp}
Suppose that Alice has neither a firm association with democrats nor with
republicans, i.e., the diagonal elements of the decision Hamiltonian are equal
to zero. Suppose also that the flipping part of the Hamiltonian contains only
the transition:%
\begin{equation}
|dr\rangle\rightarrow|rd\rangle, \label{TRCH}%
\end{equation}
which expresses the combination (democrats, republicans) into the combination
(republicans, democrats), and vice versa. Let $\lambda_{dr,rd}=\lambda
_{rd,dr}=\lambda>0.$ The Schr\"{o}dinger equation has the form of a system of
linear ordinary differential equations. The dynamics of coincidence of choices
is trivial: $i\tau\frac{dx_{dd}(t)}{dt}=0,i\tau\frac{dx_{rr}(t)}{dt}=0.$
Hence, $x_{dd}(t)=x_{dd}(0),x_{rr}(t)=x_{rr}(0).$ However, the presence of a
nontrivial transition channel, equation (\ref{TRCH}), induces fluctuations of
Alice's preferences for choices $dr$ and $rd.$ Here we have the system of two
equations:
\begin{equation}
i\tau\frac{dx_{dr}(t)}{dt}=\lambda x_{rd}(t),\;i\tau\frac{dx_{rd}(t)}%
{dt}=\lambda x_{dr}(t). \label{TRCHa}%
\end{equation}
Its solutions have the form:
\begin{equation}
x_{dr}(t)=x_{dr}(0)\cos\;\frac{\lambda t}{\tau}-ix_{rd}(0)\sin\;\frac{\lambda
t}{\tau}, \label{TRCHa1}%
\end{equation}%
\begin{equation}
x_{rd}(t)=-ix_{dr}(0)\sin\;\frac{\lambda t}{\tau}+x_{rd}(0)\cos\;\frac{\lambda
t}{\tau}. \label{TRCHa2}%
\end{equation}

\end{exmp}

\section{Dynamics in the election bath}

In physics the dynamics of a system in a bath is described by the quantum
analog of the master equation, the GKSL-equation, see section 3. We write this
equation by using the time scaling constant $\tau,$ instead of the Planck
constant:
\begin{equation}
i\tau\frac{\partial\rho}{\partial t}(t)=[H,\rho(t)]+L(\rho(t)); \label{GKSL}%
\end{equation}
where $L$ is a linear operator acting in the space of operators on the complex
Hilbert space. In the dynamics described by equation (\ref{GKSL}), density
operators are transformed into density operators. The general form of $L$ was
found by Gorini, Kossakowski, Sudarshan, and Lindblad (see, for example,
Ingarden et al. (1997) \cite{IKO}. For now, we are not interested in (the
rather complex) structure of $L.$ For our applications, it is sufficient to
know that it can be expressed through matrix multiplication for a family of
matrices. The simplest dynamics of interaction of Alice with the two party
election campaign is determined by two matrices $V_{d}$ and $V_{r}$
corresponding to advertising of democrats and republicans, respectively. Under
natural selection of the matrices $H,V_{d},V_{r}$ any solution of this
equation stabilizes to a diagonal density matrix
\begin{equation}
\rho_{\mathrm{decision}}=\mathrm{{diag}(p_{dd},p_{dr},p_{rd},p_{rr}).}
\label{DENSITY}%
\end{equation}
This matrix describes the distribution of firmly established decisions for
voting strategies $xy,$ where $x,y=d,r.$

The density matrix $\rho_{\mathrm{decision}}$ describes a population of voters
who finally determine their choices. Denote the number of people in this
population by $N.$ There are then (approximately) $n_{dd}=p_{dd}N$ people in
the mental state $|dd\rangle,$...and $n_{rr}=p_{rr}N$ people in the mental
state $|rr\rangle.$ For example, people in the mental state $|dd\rangle$
ha\textbf{ve} firmly selected to vote for democrats both in the executive and
legislative branches. Their decision is stable. From a pragmatic point of view
there is no possibility to manipulate opinions of people in this population.

\subsection{Comparing mental states given by vectors and density matrices}

Consider two populations, say $\mathcal{E}_{1}$ and $\mathcal{E}_{2}.$ Suppose
that in our QL model the first one is described by a pure state
\begin{equation}
\psi=c_{dd}|dd\rangle+c_{dr}|dr\rangle+c_{rd}|rd\rangle+c_{rr}|dd\rangle;
\label{PURE}%
\end{equation}
and the second one by the density matrix given by equation (\ref{DENSITY}).
Moreover, suppose that the complex amplitudes given by the coefficients in the
expansion (equation (\ref{PURE})) produce the same probabilities as the
density matrix, i.e.:
\begin{equation}
p_{xy}=|c_{xy}|^{2}. \label{PURE1}%
\end{equation}
One may ask: \textquotedblleft What is the difference?\textquotedblright\ At
first sight there is no difference at all, since we obtain the same
probability distribution of preferences. However, the distributions of mental
state in ensembles $\mathcal{E}_{1}$ and $\mathcal{E}_{2}$ are totally
different. All people in $\mathcal{E}_{1}$ are in the same state of
indeterminacy (superposition) $\psi.$ They are in doubt. They are ready to
change their opinion (to create a new superposition of opinions). The
$\mathcal{E}_{1}$ is a proper population for political manipulations. To the
opposite of population $\mathcal{E}_{1},$ the population $\mathcal{E}_{2}$
consists of people who have already resolved their doubts. Their mental states
have already been reduced to states of the form $xy,$ i.e. definite
choices\footnote{\textbf{Please see the discussion under equation (21).}}.

\subsection{Independence from the initial state}

The general theory of quantum master equation\textbf{s} implies that for some
important open system dynamics, the limiting probability distribution
\textit{does not depend on the initial state!} This mathematical fact has
important consequences for our QL model of elections. It tells us that in
principle it is possible to create such a quantum open system dynamics (voters
interacting with some election bath) such that the desired state
$\rho_{\mathrm{decision}}$ would be obtained \textit{independently} of the
initial mental state of Alice. This theoretical result may play an important
role in QL election technologies. Even if a quantum master equation does not
ha\textbf{ve} the unique limiting state, there are typically just a few of
them. In this case, we can split the set of all pure states (the unit sphere
in the complex Hilbert state space) into clusters of voters. For each cluster,
we can predict the final distribution of decisions.

We shall illustrate the above discussion by numerical
simulation\footnote{Using Mathematica software.} of the dynamics of
preferences of voters interacting with \textbf{the }\textquotedblleft election
environment\textquotedblright.

\begin{exmp}
We consider only the two dimensional submodel of the general four dimensional
model corresponding to a part of the electorate which have \textquotedblleft
double preferences\textquotedblright\ -- democrats in one of the elections and
republicans in another election. So, we reduce the modeling to the subspace
with the basis $|dr\rangle,|rd\rangle.$ It is assumed that at the beginning
(i.e., before interaction with the `election environment') voters are in a
superposition of the basic states:
\begin{equation}
|\psi\rangle=c_{1}|dr\rangle+c_{2}|rd\rangle,\;|c_{1}|^{2}+|c_{2}|^{2}=1.
\label{LAB}%
\end{equation}
We also assume that in the absence of interaction with the `election campaign'
the state of preferences fluctuations are driven by the Schr\"{o}dinger
dynamics considered in Example 1. In the matrix form the corresponding
Hamiltonian can be written as
\begin{equation}
\mathcal{H}=\left(
\begin{array}
[c]{ll}%
0 & \;\lambda\\
\lambda & 0
\end{array}
\right)  ; \label{BE0dj4_P}%
\end{equation}
where $\lambda>0$ is the parameter describing the intensity of flipping from
$dr$ to $rd$ and vice versa. The simplest perturbation of this Schr\"{o}dinger
equation is given by the Lindblad term of the form given by Ingarden et al.
(1997) \cite{IKO}:%
\[
C\rho C^{\ast}-(C^{\ast}C\rho+\rho C^{\ast}C)/2=C\rho C^{\ast}-\frac{1}%
{2}\{C^{\ast}C,\rho\};
\]
where $C^{\ast}$ denotes the operator which is the Hermitian adjoint to the
operator $C.$ As always in quantum formalism,
\[
\{U,V\}=UV+VU,
\]
which denotes the anticommutator of two operators$U,V.$ We select the operator
$C$ by using its matrix in the basis $|dr\rangle,|rd\rangle:$
\[
C=\left(
\begin{array}
[c]{ll}%
0 & \;\lambda\\
0 & 0
\end{array}
\right)  ;
\]
hence,
\[
C^{\ast}=\left(
\begin{array}
[c]{ll}%
0 & \;0\\
\lambda & 0
\end{array}
\right)  ;
\]
where the parameter $\lambda$ is responsible for interaction between the
voter's state. For simplicity, the `election campaign' is selected in the same
way as in the Hamiltonian (\ref{BE0dj4_P}). Thus, we proceed with the quantum
master equation:
\begin{equation}
\frac{d\rho}{dt}(t)=-i[\mathcal{H},\rho(t)]+C\rho(t)C^{\ast}-\frac{1}%
{2}\{C^{\ast}C,\rho(t)\}. \label{HHH1}%
\end{equation}
We present the dynamics corresponding to symmetric superposition,
\begin{equation}
c_{1}=c_{2}=\frac{1}{\sqrt{2}}. \label{SP}%
\end{equation}
See Fig. 1. Strongly asymmetric superposition
\begin{equation}
c_{1}=\sqrt{0,9},c_{2}=\sqrt{0.1}. \label{SPa}%
\end{equation}
See Fig. 2.
\end{exmp}

\begin{figure}[ptb]
\begin{center}
\label{FIG0505} \includegraphics[width=8cm]{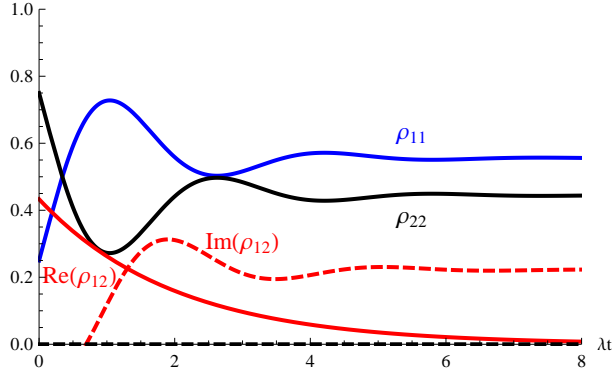}
\end{center}
\caption{Stabilization of the matrix elements of the density operator; the
initial state is symmetric superposition of state $|dr\rangle$ and
$|rd\rangle.$}%
\end{figure}

\begin{figure}[ptb]
\begin{center}
\label{FIG0901} \includegraphics[width=8cm]{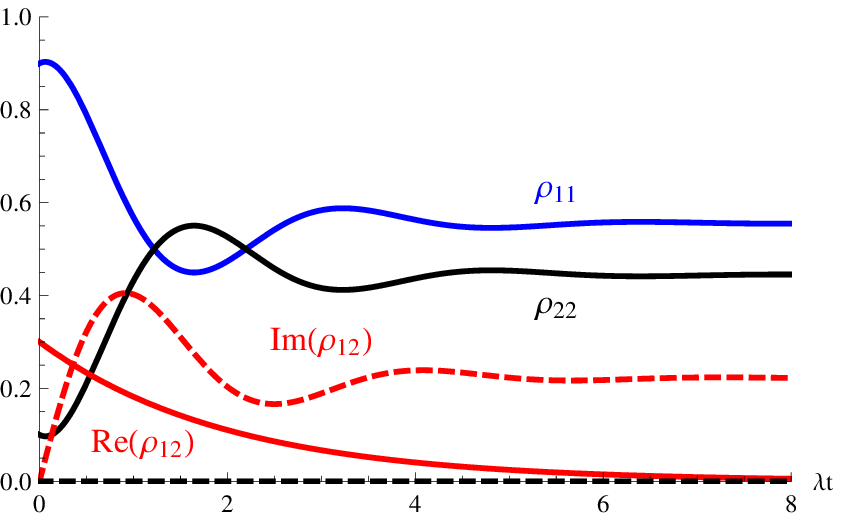}
\end{center}
\caption{Stabilization of the matrix elements of the density operator; the
initial state is stronly asymmetric superposition of states $|dr\rangle$ and
$|rd\rangle.$}%
\end{figure}

The interaction with the `election environment' plays a crucial role. Strong
oscillations of the dynamics, given by equations\textbf{\ }(\ref{TRCHa1}),
(\ref{TRCHa2}) in the absence of interaction with the `election bath' are
quickly damped and the matrix elements $\rho_{11}\equiv\rho_{dr,dr},\rho
_{22}\equiv\rho_{rd,rd},\rho_{12}\equiv\rho_{dr,rd},$ and $\bar{\rho}%
_{12}=\rho_{21}\equiv\rho_{rd,dr}$ stabilize to the definite values. Thus the
preferences of population of voters who were in fluctuating superposition of
choices stabilize under the pressure of the `election bath'. We selected such
a form of interaction between a voter and the `election bath' such that both
initial states, the totally symmetric state, i.e., no preference to $dr$ nor
$rd,$ and the state with very strong preference for the\textbf{\ }$dr$
combination in votes to congress and of president, $p(dr)=0.9,p(rd)=0.1,$
induce dynamics with stabilization to the same density matrix $\rho
_{\mathrm{lim}}.$ This example demonstrates the power of the social
environment which, in fact, determines the choices of voters.

In the $\rho_{\mathrm{lim}}$ the elements $\rho_{dr,dr}\approx0.6,\rho
_{rd,rd}\approx0.4$ determine corresponding probabilities $p(dr)\approx
0.6,p(rd)=0,4.$ Under the pressure of the social environment those who started
with a superposition as indicated in equation (\ref{SP}) increase the
$dr$-preference and those who started with the superposition in equation
(\ref{SPa}) decrease this preference, and the resulting distribution of
choices is the same in both populations (with the initial state (\ref{SP}) and
with the initial state (\ref{SPa})).

We stress that manipulation by the preferences described by the dynamics in
equation\textbf{\ }(\ref{HHH1}) in sufficiently smooth. Those dynamics are an
extension of the `free thinking' dynamics given by the Schr\"{o}dinger
equation, the first term in the right-hand side of equation\textbf{\ }%
(\ref{HHH1}). Hence, in this model the social environment does not prohibit
internal fluctuations of individuals, but instead damps them to obtain a
`peaceful' stabilization.

We emphasize that the degree of quantum uncertainty decreases in the process
of evolution. One of the standard measures of uncertainty which is used in
quantum information theory is given by so called \textit{linear entropy} (see
Ingarden et al. (1997) \cite{IKO}) defined as:%
\[
S_{L}=1-\mathrm{{Tr}\rho^{2}.}%
\]
For a pure state (which has the highest degree of uncertainty), the linear
entropy $S_{L,\mathrm{min}}=0.$ It increases with degeneration of purity in a
quantum state and it approaches it maximal value $S_{L,\mathrm{max}}=0,5$ for
\textbf{a }maximally mixed state. Here we consider the two dimensional case;
in the general case $S_{L,\mathrm{max}}=1-1/d,$ where $d$ is the dimension of
the state space. The dynamics of linear entropy corresponding to the initial
states as per equations (\ref{SP}) and (28), respectively, are presented
\textbf{in} Fig. 3 and Fig 4.

\begin{figure}[ptb]
\begin{center}
\label{FIG_E0505} \includegraphics[width=8cm]{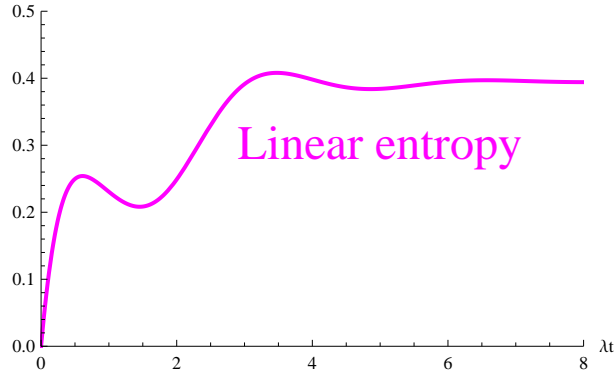}
\end{center}
\caption{Stabilization of linear entropy: the initial state is symmetric
superposition of states $|dr\rangle$ and $|rd\rangle.$}%
\end{figure}

\begin{figure}[ptb]
\begin{center}
\label{FIG_E0901} \includegraphics[width=8cm]{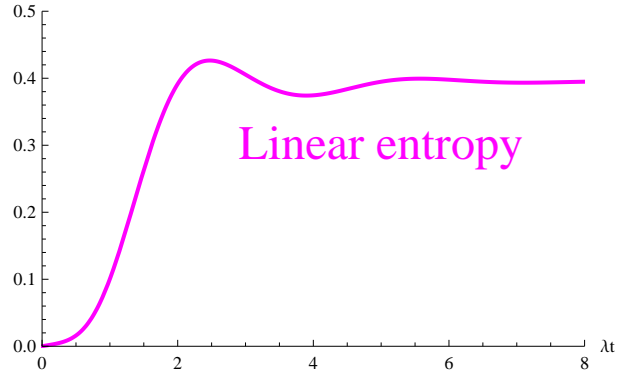}
\end{center}
\caption{Stabilization of of linear entropy: the initial state is strongly
asymmetric superposition of states $|dr\rangle$ and $|rd\rangle.$}%
\end{figure}

We see that \textbf{the }entropy behave\textbf{s} in different ways, but
finally it stabilizes to the same value $S_{L}\approx0.4.$ \textbf{T}his value
corresponds to a very large decreasing of purity -- uncertainty of the
superposition type.

Numerical simulation demonstrated that, for other choices of pure initial
states, the density matrix and the linear entropy stabilize to the same
values. Our conjecture is that it may be possible to prove theoretically that
this is really the case. However, at the moment we have only results of
numerical simulation supporting this conjecture.

\end{document}